\documentclass[12pt]{iopart}
\usepackage{iopams}
\usepackage{graphicx}
\usepackage[dvips]{color}
\usepackage{amssymb, amsfonts, amsbsy}

\begin{document}
\title[The Hubble series: Convergence properties and redshift variables]    
{The Hubble series: \\ Convergence properties and redshift variables}
\author{C\'eline Catto\"en and Matt Visser}
\address{School of Mathematics, Statistics, and Computer Science, \\
Victoria University of Wellington, PO Box 600, Wellington, New Zealand}
\ead{celine.cattoen@mcs.vuw.ac.nz, matt.visser@mcs.vuw.ac.nz}
\begin{abstract}

In cosmography, cosmokinetics, and cosmology it is quite common to encounter physical quantities expanded as a Taylor series in the cosmological redshift $z$. Perhaps the most well-known exemplar of this phenomenon is the Hubble relation between distance and redshift. 
However, we now have considerable high-$z$ data available, for instance we have supernova data  at least back to redshift $z\approx 1.75$. This opens up the theoretical question as to whether or not the Hubble series (or more generally any series expansion based on the $z$-redshift) actually converges for large redshift?
Based on a combination of mathematical and physical reasoning, we argue that the radius of convergence of any series expansion in $z$ is less than or equal to 1, and that $z$-based expansions must break down for $z>1$, corresponding to a universe less than half its current size.   
  
Furthermore, we shall argue on theoretical grounds for the utility of an
  improved parameterization $y=z/(1+z)$.  In terms of the $y$-redshift we again argue that the radius of convergence of any series expansion in $y$ is less than or equal to 1, so that $y$-based expansions are  likely to be good all the way back to the big bang  ($y=1$), but that $y$-based expansions must break down for $y<-1$, now corresponding to a universe more than twice its current size.   

\vskip 0.250cm


\vskip 0.1250cm
\noindent
 arXiv:  30 July 2007; 
\LaTeX-ed \today.
  
\end{abstract}
\maketitle
\newtheorem{theorem}{Theorem}
\newtheorem{corollary}{Corollary}
\newtheorem{lemma}{Lemma}
\def\d{{\mathrm{d}}}
\def\implies{\Rightarrow}

\def\eg{{\it e.g.}}
\def\ie{{\it i.e.}}
\def\etc{{\it etc.}}
\def\sign{{\hbox{sign}}}
\def\eof{\Box}
\newenvironment{warning}{{\noindent\bf Warning: }}{\hfill $\eof$\break}
\markboth{The Hubble series: Convergence properties and redshift variables
}{}
\clearpage
\section{Introduction}

Consider the standard luminosity distance versus redshift relation~\cite{Weinberg, Peebles}:
\begin{eqnarray}
d_L(z) =  {c\; z\over H_0}
\Bigg\{ 1 + {1\over2}\left[1-q_0\right] {z} 
+ O(z^2) \Bigg\},
\label{E:Hubble1a}
\end{eqnarray}
and its higher-order extension~\cite{Chiba, Sahni, Jerk, Jerk2}
\begin{eqnarray}
\fl
d_L(z) =  {c\; z\over H_0}
\Bigg\{ 1 + {1\over2}\left[1-q_0\right] {z} 
-{1\over6}\left[1-q_0-3q_0^2+j_0+ {kc^2\over H_0^2\,a_0^2}\right] z^2
+ O(z^3) \Bigg\}.
\label{E:Hubble1}
\end{eqnarray}
As will quickly be verified by looking at the derivation (see, for example,~\cite{Weinberg, Peebles, Chiba, Sahni, Jerk, Jerk2}, the standard Hubble law is actually a Taylor series expansion derived for small $z$, whereas much of the most interesting recent supernova data occurs at $z\gtrsim1$~\cite{legacy, legacy-url, gold, Riess2006a, Riess2006b, essence}. Should we even trust the usual formalism for large $z>1$?
Two distinct things could go wrong~\cite{Hubble-arXiv}:
\begin{itemize}
\item The underlying Taylor series could fail to converge.
\item Finite truncations of the Taylor series might be a
               bad approximation to the exact result.
\end{itemize}
In fact, \emph{both} things happen. There are good mathematical and physical reasons for this undesirable behaviour, as we shall discuss below. We shall carefully explain just what goes wrong --- and suggest various ways of improving the situation. Our ultimate goal will be to find suitable forms of the Hubble relation that are well adapted to performing fits to all the available distance \emph{versus} redshift data. More generally, the same sort of argument applies to any physical quantity that is expanded as a Taylor series in the redshift. Based on a combination of mathematical and physical reasoning, we argue that the radius of convergence of any series expansion in $z$ is less than or equal to 1, and that $z$-based expansions must break down for $z>1$, corresponding to a universe less than half its current size.   Furthermore, $z$-based expansions are  likely to be good all the way to $z=-1$, corresponding to a universe that has grown to infinite size.
  
We shall then argue on theoretical grounds for the utility of an improved parameterization $y=z/(1+z)$.  In terms of the $y$-redshift we again argue that the radius of convergence of any series expansion in $y$ is less than or equal to 1, so that $y$-based expansions are  likely to be good all the way back to the big bang ($y=1$), but that $y$-based expansions must break down for $y<-1$, now corresponding to a universe more than twice its current size.   
Choosing to adopt the $z$-redshift or the $y$-redshift  will not change the physics, \emph{but it might improve mathematical convergence and physical insight}. 

\section{Cosmological distance scales}

In cosmology there are many different and equally natural definitions of the notion of   ``distance'' between two objects or events, whether directly observable or not.
For the vertical axis of the Hubble plot, instead of using the standard default choice of luminosity distance $d_L$, let  us now consider using one or more of:
\begin{itemize}

\item The ``photon flux distance'':
\begin{equation}
 d_F = {d_L\over(1+z)^{1/2}} .
\end{equation} 

\item The ``photon count distance'':
\begin{equation}
 d_P = {d_L\over(1+z)}.
\end{equation} 

\item The  ``deceleration distance'':
\begin{equation}
 d_Q = {d_L\over(1+z)^{3/2}}.
\end{equation} 

\item  The ``angular diameter distance'':
\begin{equation}
 d_A = {d_L\over(1+z)^2}.
\end{equation} 

\item The ``distance modulus'':
\begin{equation}
\mu_D = {5} \; \log_{10}[d_L/(10 \hbox{ pc})] = {5} \; \log_{10}[d_L/(1 \hbox{ Mpc})] +25.
\end{equation}

\item Or possibly some other surrogate for distance.
\end{itemize}
Some words of explanation and caution are in order here~\cite{Hubble-arXiv, Hogg}: 
\begin{itemize}

\item
The ``photon flux distance'' $d_F$ is based on the fact that it is often technologically easier to count the photon flux (photons/sec) than it is to bolometrically measure total energy flux (power) deposited in the detector. If we are counting photon number flux, rather than energy flux, then the photon number flux contains one fewer factor of $(1+z)^{-1}$. Converted to a distance estimator, the ``photon flux distance'' contains one extra factor of $(1+z)^{-1/2}$ as compared to the (power-based) luminosity distance.

\item 
The ``photon count distance''  $d_P$  is related to the total number of photons absorbed without regard to the rate at which they arrive. Thus the ``photon count distance'' contains one extra factor of $(1+z)^{-1}$ as compared to the (power-based) luminosity distance. Indeed D'Inverno~\cite{dInverno} uses what is effectively this photon count distance as his nonstandard definition for luminosity distance. Furthermore, though motivated very differently, this quantity is equal to Weinberg's definition of proper motion distance~\cite{Weinberg}, and is also equal to Peebles' version of angular diameter distance~\cite{Peebles}.  That is:
\begin{equation}
d_P= d_{L,\hbox{\scriptsize D'Inverno}}  = d_{\mathrm{proper},\mathrm{Weinberg}} = d_{A,\mathrm{Peebles}} .
\end{equation} 

\item 
The quantity $d_Q$ is (as far as we can tell)  a previously un-named quantity that seems to have no simple direct physical interpretation --- but we shall soon see why it is potentially useful, and why it is useful to refer to it as the ``deceleration distance''.

\item 
The quantity $d_A$ is Weinberg's definition of angular diameter distance~\cite{Weinberg}, corresponding to the physical size of the object \emph{when the light was emitted}, divided by its current angular diameter on the sky. This differs from Peebles' definition of angular diameter distance~\cite{Peebles}, which corresponds to what the size of the object would be at the current cosmological epoch if it had continued to co-move with the cosmological expansion (that is, the ``comoving size''), divided by its current angular diameter on the sky. Weinberg's $d_A$ exhibits the (at first sight perplexing, but physically correct) feature that beyond a certain point $d_A$ can actually \emph{decrease} as one moves to older objects that are clearly ``further'' away. In contrast Peebles' version of angular diameter distance is always increasing as one moves ``further'' away. Note that
\begin{equation}
d_{A,\mathrm{Peebles}}  = (1+z)\; d_A.
\end{equation}

\item 
Finally, note that the distance modulus can be rewritten in terms of traditional stellar magnitudes as 
\begin{equation}
\mu_D = \mu_\mathrm{apparent} - \mu_\mathrm{absolute}.
\end{equation}
The continued use of stellar magnitudes and the distance modulus in the context of cosmology is largely a matter of historical tradition, though we shall soon see that the logarithmic nature of the distance modulus has interesting and useful side effects. Note that we prefer as much as possible to deal with natural logarithms: $\ln x = \ln(10) \; \log_{10} x$. Indeed
\begin{equation}
\label{E:mu}
\mu_D = {5\over\ln10} \; \ln[d_L/(1 \hbox{ Mpc})] +25,
\end{equation}
so that
\begin{equation}
\label{E:d}
\ln[d_L/(1 \hbox{ Mpc})] = {\ln10\over5} [\mu_D - 25].
\end{equation}
\end{itemize}
Obviously
\begin{equation}
d_L \geq d_F \geq d_P \geq d_Q \geq d_A.
\end{equation} 
Furthermore these particular distance scales satisfy the property that they converge on each other, and converge on the naive Euclidean notion of distance, as $z\to0$.

To simplify subsequent formulae, it is now useful to define  the  ``Hubble distance''~\footnote{The ``Hubble distance'' $d_H = c/H_0$  is sometimes called the ``Hubble radius'', or the ``Hubble sphere", or even the ``speed of light sphere" [SLS]~\cite{Rothman}. Sometimes ``Hubble distance'' is used to refer to the naive estimate $d = d_H \; z$ coming from the linear part of the Hubble relation and ignoring all higher-order terms --- this is definitely \emph{not} our intended meaning.}
\begin{equation}
d_H = {c\over H_0},
\end{equation}
so that for $H_0 = 73\, {+ 3\atop- 4} \hbox{ (km/sec)/Mpc}$~\cite{PDG} we have
\begin{equation}
d_H = 4100\, {\textstyle {+ 240\atop- 160}} \hbox{ Mpc}.
\end{equation}
Furthermore we choose to set
\begin{equation}
\Omega_0=1+{kc^2\over H_0^2a_0^2} = 1 + {k \; d_H^2\over a_0^2}.
\end{equation}
For our purposes $\Omega_0$ is a purely cosmographic definition without dynamical content. (Only if one additionally invokes the Einstein equations in the form of the Friedmann equations does $\Omega_0$ have the standard interpretation as the ratio of total density to the Hubble density, but we would be prejudging things by making such an identification in the current cosmographic framework.) In the cosmographic framework $k/a_0^2$ is simply the present day curvature of space (not spacetime), while $d_H^{\;-2}=H_0^2/c^2$ is a measure of the contribution of expansion to the spacetime curvature of the FLRW geometry. More precisely, in a FRLW universe the Riemann tensor has (up to symmetry) only two non-trivial components. In an orthonormal basis:
\begin{equation}
R_{\hat\theta\hat\phi\hat\theta\hat\phi} = {k\over a^2} + {\dot a^2\over c^2 \; a^2}
= {k\over a^2} + {H^2\over c^2};
\end{equation}
\begin{equation}
R_{\hat t\hat r\hat t\hat r} = - {\ddot a\over c^2\; a } =  {q \; H^2\over c^2}.
\end{equation}
Then at arbitrary times $\Omega$ can be defined purely in terms of the Riemann tensor of the FLRW spacetime as   
\begin{equation}
\Omega = 1 + { R_{\hat\theta\hat\phi\hat\theta\hat\phi}(\dot a\to 0) 
           \over R_{\hat\theta\hat\phi\hat\theta\hat\phi} (k\to 0)}.
\end{equation}

\section{New versions of the Hubble law}

New versions of the Hubble law are easily calculated for each of these cosmological distance scales. 
Explicitly:
\begin{eqnarray}
\fl
d_L(z) =  {d_H \; z }
\Bigg\{ 1 - {1\over2}\left[-1+q_0\right] {z} 
+{1\over6}\left[q_0+3q_0^2-(j_0+\Omega_0)\right] z^2
+ O(z^3) \Bigg\}.
\end{eqnarray}

\begin{eqnarray}
\fl
d_F(z) =  {d_H \; z }
\Bigg\{ 1 - {1\over2}q_0 {z} 
+{1\over24}\left[3+10q_0+12q_0^2-4(j_0+\Omega_0)\right] z^2
+ O(z^3) \Bigg\}.
\end{eqnarray}

\begin{eqnarray}
\fl
d_P(z) =  {d_H \; z }
\Bigg\{ 1 - {1\over2}\left[1+q_0\right] {z} 
+{1\over6}\left[3+4q_0+3q_0^2-(j_0+\Omega_0)\right] z^2
+ O(z^3) \Bigg\}.
\end{eqnarray}

\begin{eqnarray}
\fl
d_Q(z) =  {d_H \; z }
\Bigg\{ 1   - {1\over2}\left[2+q_0\right] {z}
+{1\over24}\left[27+22q_0+12q_0^2-4(j_0+\Omega_0)\right] z^2
+ O(z^3) \Bigg\}.
\end{eqnarray}

\begin{eqnarray}
\fl
d_A(z) =  {d_H \; z }
\Bigg\{ 1 - {1\over2}\left[3+q_0\right] {z} 
+{1\over6}\left[12+7q_0+3q_0^2-(j_0+\Omega_0)\right] z^2
+ O(z^3) \Bigg\}.
\end{eqnarray}
If one simply wants to deduce (for instance)  the \emph{sign} of $q_0$, then it seems that plotting the ``photon flux distance'' $d_F$ versus $z$ would be a particularly good test --- simply check if the first nonlinear term in the Hubble relation curves up or down. 

In contrast, the Hubble law for the distance modulus itself is given by the more complicated expression
\begin{eqnarray}
\fl
\mu_D(z) &=&  25 + {5\over\ln(10)} \Bigg\{ \ln(d_H/\hbox{Mpc}) + \ln z 
\nonumber\\
\fl
&&
 + {1\over2}\left[1-q_0\right] {z} 
-{1\over24}\left[3-10q_0-9q_0^2+4(j_0+\Omega_0)\right] z^2
+ O(z^3) \Bigg\} .
\end{eqnarray}
However, when plotting $\mu_D$ versus $z$, most of the observed curvature in the plot comes from the universal ($\ln z$) term, and so carries no real information and is relatively uninteresting.  It is much better to rearrange the above as:
\begin{eqnarray}
\fl
\ln[d_L/(z \hbox{ Mpc})] &=& {\ln10\over5} [\mu_D - 25] - \ln z
\nonumber\\
\fl
&=& \ln(d_H/\hbox{Mpc})
\nonumber\\
\fl
&&
 - {1\over2}\left[-1+q_0\right] {z} 
+{1\over24}\left[-3+10q_0+9q_0^2-4(j_0+\Omega_0)\right] z^2
+ O(z^3).
\end{eqnarray}
In a similar manner one has
\begin{eqnarray}
\fl
\ln[d_F/(z \hbox{ Mpc})] &=& {\ln10\over5} [\mu_D - 25] - \ln z - {1\over2} \ln(1+z)
\nonumber\\
\fl
&=& \ln(d_H/\hbox{Mpc})
\nonumber\\
\fl
&&
 - {1\over2}q_0  {z} 
+{1\over24}\left[3+10q_0+9q_0^2-4(j_0+\Omega_0)\right] z^2
+ O(z^3).
\end{eqnarray}

\begin{eqnarray}
\fl
\ln[d_P/(z \hbox{ Mpc})] &=& {\ln10\over5} [\mu_D - 25] - \ln z -  \ln(1+z)
\nonumber\\
\fl
&=& \ln(d_H/\hbox{Mpc})
\nonumber\\
\fl
&&
 - {1\over2}\left[1+q_0\right]  {z} 
+{1\over24}\left[9+10q_0+9q_0^2-4(j_0+\Omega_0)\right] z^2
+ O(z^3).
\end{eqnarray}

\begin{eqnarray}
\fl
\ln[d_Q/(z \hbox{ Mpc})] &=& {\ln10\over5} [\mu_D - 25] - \ln z - {3\over2} \ln(1+z)
\nonumber\\
\fl
&=& \ln(d_H/\hbox{Mpc})
\nonumber\\
\fl
&&
 - {1\over2}\left[2+q_0\right]  {z} 
+{1\over24}\left[15+10q_0+9q_0^2-4(j_0+\Omega_0)\right] z^2
+ O(z^3).
\end{eqnarray}

\begin{eqnarray}
\fl
\ln[d_A/(z \hbox{ Mpc})] &=& {\ln10\over5} [\mu_D - 25] - \ln z - 2 \ln(1+z)
\nonumber\\
\fl
&=& \ln(d_H/\hbox{Mpc})
\nonumber\\
\fl
&&
 - {1\over2}\left[ 3 + q_0\right]  {z} 
+{1\over24}\left[21+10q_0+9q_0^2-4(j_0+\Omega_0)\right] z^2
+ O(z^3).
\end{eqnarray}
These logarithmic versions of the Hubble law have several advantages --- fits to these relations are easily calculated in terms of the observationally reported distance moduli $\mu_D$ and their estimated statistical uncertainties~\cite{legacy,  legacy-url, gold, Riess2006a, Riess2006b}.  (Specifically there is no need to transform the statistical uncertainties on the distance moduli beyond a universal multiplication by the factor $[\ln 10]/5$.) Furthermore the deceleration parameter $q_0$ is easy to extract as it has been ``untangled'' from both Hubble parameter and the combination ($j_0+\Omega_0$).

Note that it is always the combination ($j_0+\Omega_0$) that arises in these third-order Hubble relations, and that it is even in principle impossible to separately determine $j_0$ and $\Omega_0$ in a cosmographic framework.  The reason for this degeneracy is (or should be) well-known~\cite[p. 451]{Weinberg}: Consider the \emph{exact} expression for the luminosity distance in any FLRW universe, which is usually presented in the form~\cite{Weinberg, Peebles}
\begin{equation}
\label{E:exact}
d_L(z) = a_0 \; (1+z) \;
\sin_k \left\{ {c\over H_0 \, a_0} \int_0^z {H_0\over H(z)} \; \d z \right\},
\end{equation} 
where
\begin{equation}
\sin_k(x) = 
\left\{ 
\begin{array}{ll}
       \sin(x), & k=+1;\\
        x,       & k=0;\\
        \sinh(x), & k=-1.\\
\end{array} \right.
\end{equation} 
By inspection, even if one knows $H(z)$ exactly for all $z$ one cannot determine $d_L(z)$ without independent knowledge of $k$ and $a_0$. Conversely even if one knows $d_L(z)$ exactly for all $z$ one cannot determine $H(z)$ without independent knowledge of $k$ and $a_0$. Indeed let us rewrite this exact result in a slightly different fashion as
\begin{equation}
d_L(z) = a_0 \; (1+z) \; 
{\sin\left\{ {\displaystyle {\sqrt{k} \, d_H\over a_0} \; \int_0^z {H_0\over H(z)}\;\d z } \right\} 
\over 
\sqrt{k} },
\end{equation}
where this result now holds for all $k$ provided we interpret the $k=0$ case in the obvious limiting fashion.
Equivalently, using the cosmographic $\Omega_0$ as defined above we have the \emph{exact} cosmographic result that for all $\Omega_0$:
\begin{equation}
d_L(z) = d_H \; (1+z) \; {\sin\left\{ \sqrt{\Omega_0-1} \;
{\displaystyle \int_0^z  {H_0\over H(z)}} \;\d z \right\} \over \sqrt{\Omega_0-1} }.
\end{equation}
This form of the exact Hubble relation makes it clear that an independent determination of $\Omega_0$ (equivalently, $k/a_0^2$), is needed to complete the link between $a(t)$ and $d_L(z)$. When Taylor expanded in terms of $z$, this expression leads to a degeneracy at third-order, which is  where $\Omega_0$ [equivalently $k/a_0^2$] first enters into the Hubble series~\cite{Jerk, Jerk2}.

What message should we take from this discussion? There are many physically equivalent versions of the Hubble law, corresponding to many slightly different physically reasonable definitions of distance, and whether we choose to present the Hubble law linearly or logarithmically.  If one were to have arbitrarily small scatter/error bars on the observational data, then the choice of which Hubble law one chooses to fit to would not matter. In the presence of significant scatter/uncertainty there is a risk that the fit might depend strongly on the choice of Hubble law one chooses to work with. (And if the resulting values of the deceleration parameter one obtains do depend significantly on which distance scale one uses, this is evidence that one should be very cautious in interpreting the results.)
Note that the two versions of the Hubble law based on ``photon flux distance''  $d_F$ stand out in terms of making the deceleration parameter  easy to visualize and extract.

\section{Why is the redshift expansion badly behaved for $z>1$?}

In addition to the question of which distance measure one chooses to use, there is a basic and fundamental physical and mathematical reason why the traditional redshift expansion breaks down for $z>1$.

\subsection{Convergence}

Consider the exact Hubble relation (\ref{E:exact}). 
This is certainly nicely behaved, and possesses no obvious poles or singularities, (except possibly at a  turnaround event where $H(z)\to0$, more on this below).
However if we attempt to develop a Taylor series expansion in redshift $z$, using what amounts to the \emph{definition} of the Hubble $H_0$, deceleration $q_0$,
 and jerk $j_0$ parameters, then:
\begin{equation}
\fl
{1\over 1+z} =  {a(t)\over a_0} = 1 + H_0 \; (t-t_0) -  {q_0 \; H_0^2\over 2!}  \;(t-t_0)^2 
+{j_0\; H_0^3\over 3!} \;(t-t_0)^3
+ O([t-t_0]^4).
\end{equation} 
Now this particular Taylor expansion manifestly has a pole at $z=-1$, corresponding to the instant (either  at finite or infinite time) when the universe has expanded to infinite volume, $a=\infty$. Note that a \emph{negative} value for $z$  corresponds to $a(t)>a_0$, that is: In an expanding universe $z<0$ corresponds to the \emph{future}.
Since there is an explicit pole at $z=-1$, by standard complex variable theory the radius of convergence is \emph{at most} $|z|=1$, so that this series \emph{also} fails to converge for $z > 1$, when the universe was less than half its current size. 

Consequently when reverting this power series to obtain lookback time $T=t_0-t$ as a function $T(z)$ of $z$, we should not expect that series to converge for $z >1$.  Ultimately, when written in terms of $a_0$, $H_0$, $q_0$, $j_0$, and a power series expansion in redshift $z$ you should not expect $d_L(z)$ to converge for $z > 1$.

\begin{figure}[htb]
\begin{center}
\input{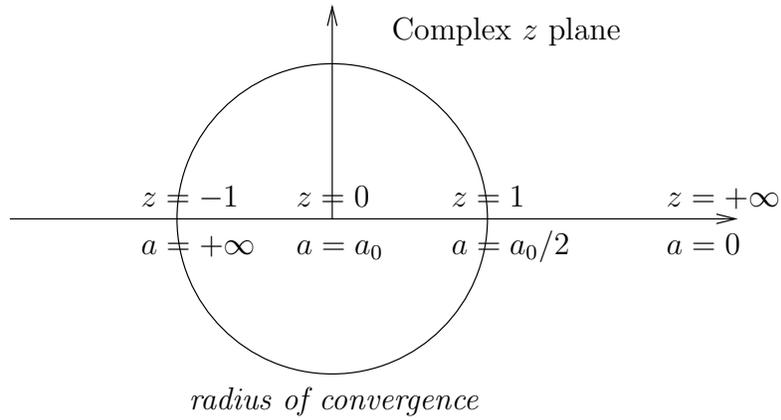}
\end{center}
\caption{\label{F:convergence_radius_z}
Qualitative sketch of the behaviour of the scale factor $a$ and the radius of convergence of the Taylor series in $z$-redshift.

}
\end{figure}

Note that the \emph{mathematics} that goes into this result is that the radius of convergence of any power series is the distance to the closest singularity in the complex plane, while the relevant \emph{physics} lies in the fact that on physical grounds we should not expect to be able to extrapolate forwards  beyond $a=\infty$, corresponding to $z=-1$. Physically we should expect this argument to hold for any observable quantity when expressed as a function of redshift and Taylor expanded around $z=0$ --- the radius of convergence of the Taylor series must be less than or equal to unity. (Note that the radius of convergence might actually be \emph{less} than unity, this occurs if some other singularity in the complex $z$ plane is closer than the breakdown in predictability associated with attempting to drive $a(t)$ ``past'' infinite expansion, $a=\infty$.) Figure \ref{F:convergence_radius_z} illustrates the radius of convergence in the complex plane of the Taylor series expansion in terms of $z$.

Consequently, we must conclude that observational data regarding $d_L(z)$ for $z > 1$ is not going to be particularly useful in fitting $a_0$, $H_0$, $q_0$, and $j_0$, to the usual \emph{traditional} version of the Hubble relation.

\subsection{Pivoting}
A trick that is sometimes used to improve the behaviour of the Hubble law is to Taylor expand around some nonzero value of $z$, which might be called the ``pivot''. That is, we take
\begin{equation}
 z = z_{pivot}+\Delta z,
\end{equation} 
and expand in powers of $\Delta z$. If we choose to do so, then observe
\begin{equation}
\fl
{1\over 1+z_{pivot}+\Delta z} =  1 + H_0 \; (t-t_0) - {1\over2} \; q_0 \; H_0^2 \;(t-t_0)^2 
+{1\over3!}\;  j_0\; H_0^3 \;(t-t_0)^3 
+ O([t-t_0]^4).\;\;
\end{equation} 
The pole is now located at:
\begin{equation}
\Delta z = -(1+z_{pivot}),
\end{equation} 
which again physically corresponds to a universe that has undergone infinite expansion, $a=\infty$. The radius of convergence is now 
\begin{equation}
|\Delta z| \leq (1+z_{pivot}),
\end{equation}
and we expect the pivoted version of the Hubble law to fail for
\begin{equation}
z > 1 + 2 \; z_{pivot}.
\end{equation} 
So pivoting is certainly helpful, and can in principle extend the convergent region of the Taylor expanded Hubble relation to somewhat higher values of $z$, but maybe we can do even better?

\subsection{Other singularities}
Other singularities that might further restrict the radius of convergence of  the Taylor expanded Hubble law (or any other Taylor expanded physical observable) are also important. Chief among them are the singularities (in the Taylor expansion) induced by turnaround events. If the universe has a minimum scale factor $a_\mathrm{min}$ (corresponding to a ``bounce'') then clearly it is meaningless to expand beyond
\begin{equation}
1+z_\mathrm{max} = a_0/a_\mathrm{min};  \qquad z_\mathrm{max} = a_0/a_\mathrm{min}-1;
\end{equation}
implying that we should restrict our attention to the region
\begin{equation}
|z| <  z_\mathrm{max} = a_0/a_\mathrm{min}-1.
\end{equation}
Since for other reasons we had already decided we should restrict attention to $|z|<1$, and since on observational grounds we certainly expect any ``bounce'', if it occurs at all, to occur for $z_\mathrm{max}\gg 1$, this condition provides no new information. 

On the other hand, if the universe has a moment of maximum expansion, and then begins to recollapse, then it is meaningless to extrapolate beyond
\begin{equation}
1+z_\mathrm{min} = a_0/a_\mathrm{max};  \qquad z_\mathrm{min} = -[1-a_0/a_\mathrm{max}];
\end{equation}
implying that we should restrict our attention to the region
\begin{equation}
|z| <  1 - a_0/a_\mathrm{max}.
\end{equation}
This relation now does provide us with additional constraint, though (compared to the $|z|<1$ condition) the bound is not appreciably tighter unless we are ``close" to a point of maximum expansion. Other singularities could lead to additional constraints.

\section{Improved redshift variable for the Hubble relation}

Now it must be admitted that the traditional redshift has a particularly simple physical interpretation:
\begin{equation}
 1+z = {\lambda_0\over\lambda_e} = {a(t_0)\over a(t_e)},
\end{equation} 
so that
\begin{equation}
z = {\lambda_0-\lambda_e\over\lambda_e} = {\Delta\lambda\over \lambda_e}.
\end{equation} 
That is, $z$ is the change in wavelength divided by the \emph{emitted} wavelength.
This is certainly simple, but there's at least one other \emph{equally simple} choice. Why not use:
\begin{equation}
y = {\lambda_0-\lambda_e\over\lambda_0} = {\Delta\lambda\over \lambda_0}\;?
\end{equation} 
That is, define $y$ to be the change in wavelength divided by the \emph{observed} wavelength.
This implies
\begin{equation}
 1-y = {\lambda_e\over\lambda_0} = {a(t_e)\over a(t_0)} = {1\over1+z}.
\end{equation} 
Now similar expansion variables have certainly been considered before. (See, for example,  Chevalier and   Polarski~\cite{Polarski}, who effectively worked with the dimensionless quantity $b=a(t)/a_0$, so that $y=1-b$. Similar ideas have also appeared in several related works~\cite{Linder, Linder2, Bassett, Martin}. Note that these authors have typically been interested in parameterizing the so-called $w$-parameter, rather than specifically addressing the Hubble relation.)

Indeed, the variable $y$ introduced above has some very nice properties:
\begin{equation}
y = {z\over1+z}; \qquad z={y\over1-y}.
\end{equation} 
In the past (of an expanding universe)
\begin{equation}
z \in (0,\infty); \qquad  y\in (0,1);
\end{equation}
while in the future
\begin{equation}
z \in (-1,0); \qquad  y\in (-\infty,0).
\end{equation}
So the variable $y$ is both easy to compute, and when extrapolating back to the Big Bang has a nice finite range $(0,1)$.
We will refer to this variable as the \emph{$y$-redshift}.
(Originally when developing these ideas we had intended to use the variable $y$ to develop orthogonal polynomial expansions on the finite interval $y\in [0,1]$. This is certainly possible, but we shall soon see that given the current data, this is somewhat overkill, and simple polynomial fits in $y$ are adequate for our purposes.)

In terms of the variable $y$ it is easy to extract a new version of the Hubble law by simple substitution:
\begin{eqnarray} \label{dL}
\fl
d_L(y) =  {d_H\; y}
\Bigg\{ 1 - {1\over2}\left[-3+q_0\right] {y} 
+{1\over6}\left[12-5q_0+3q_0^2-(j_0+ \Omega_0) \right] y^2
+ O(y^3) \Bigg\}.
\end{eqnarray}
This still looks rather messy, in fact as messy as before  --- one might justifiably ask in what sense is this new variable any real improvement?

\begin{figure}[htb]
\begin{center}
\input{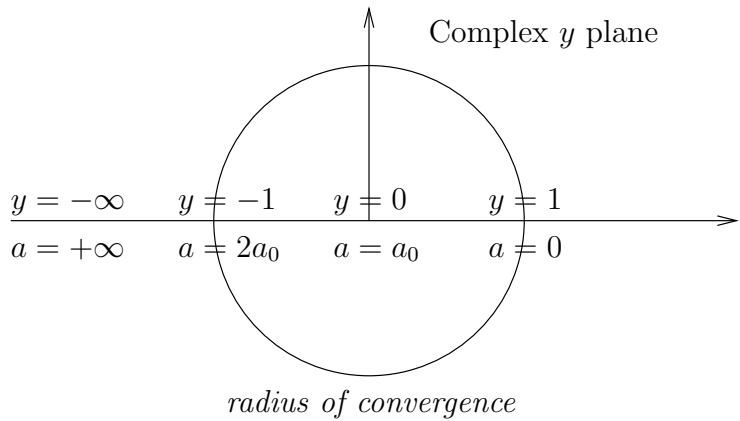}
\end{center}
\caption{\label{F:convergence_radius_y}
Qualitative sketch of the behaviour of the scale factor $a$ and the radius of convergence of the Taylor series in $y$-redshift.
}
\end{figure}

First, when expanded in terms of $y$, the formal radius of convergence covers much more of the physically interesting region. Consider:
\begin{eqnarray}
\fl
1-y =  1 + H_0 \; (t-t_0) - {1\over2} \; q_0 \; H_0^2 \;(t-t_0)^2 
+{1\over3!}\;  j_0\; H_0^3 \;(t-t_0)^3
+ O([t-t_0]^4).
\end{eqnarray}
This expression now has no poles, so upon reversion of the series lookback time $T=t_0-t$ should be well behaved as a function $T(y)$ of $y$ --- at least all the way back to the Big Bang.
(We now expect, on physical grounds, that the power series is likely to break down if one tries to extrapolate backwards \emph{through} the Big Bang.) Based on this, we now expect $d_L(y)$, as long as it is expressed as a Taylor series in the variable $y$, to be a well-behaved power series
all the way to the Big Bang. 
In fact, since
\begin{equation}
 y = +1    \qquad \Leftrightarrow  \qquad    \hbox{Big Bang},
\end{equation} 
we expect the radius of convergence to be given by $|y|=1$, so that the series converges for 
\begin{equation}
 |y| < 1.
\end{equation} 
Consequently, when looking into the future, in terms of the variable $y$ we expect to encounter problems at $y = -1$, when the universe
    has expanded to twice its current size.  Figure \ref{F:convergence_radius_y} illustrates the radius of convergence in the complex plane of the Taylor series expansion in terms of $y$.

Note the tradeoff here --- $z$ is a useful expansion parameter for arbitrarily large universes, but breaks down for a universe half its current size or less; in contrast $y$ is a useful expansion parameter all the way back to the Big Bang, but breaks down for a universe double its current size or more. Whether or not $y$ is more suitable than $z$ depends very much on what you are interested in doing. This is illustrated in Figures \ref{F:convergence_radius_z} and \ref{F:convergence_radius_y}. For the purposes of this article we are interested in high-redshift supernovae --- and we want to probe rather early times --- so it is definitely $y$ that is more appropriate here. Indeed the furthest supernova for which we presently have both spectroscopic data and an estimate of the distance occurs at $z=1.755$~\cite{Riess2006a},  corresponding to $y=0.6370$.
Furthermore, using the variable $y$ it is easier to plot very large redshift datapoints. For example, (though we shall not pursue this point in this article), the Cosmological Microwave Background is located at $z_\mathrm{CMB}=1088$, which corresponds to $y_\mathrm{CMB}=0.999$. This point is not ``out of range'' as it would be if one uses the variable $z$.

\section{More versions of the Hubble law}

In terms of this new redshift variable, the ``linear in distance'' Hubble relations are:
\begin{eqnarray}
\fl
d_L(y) =  {d_H\; y}
\Bigg\{ 1 - {1\over2}\left[-3+q_0\right] {y} 
+{1\over6}\left[12-5q_0+3q_0^2-(j_0+\Omega_0) \right] y^2
+ O(y^3) \Bigg\}.
\end{eqnarray}

\begin{eqnarray}
\fl
d_F(y) =  {d_H\; y}
\Bigg\{ 1 - {1\over2}\left[-2+q_0\right] {y} 
+{1\over24}\left[27-14q_0+12q_0^2-4(j_0+\Omega_0)\right] y^2
+ O(y^3) \Bigg\}.
\end{eqnarray}

\begin{eqnarray}
\fl
d_P(y) =  {d_H\; y}
\Bigg\{ 1 - {1\over2}\left[-1+q_0\right] {y} 
+{1\over6}\left[3-2q_0+3q_0^2-(j_0+ \Omega_0)\right] y^2
+ O(y^3) \Bigg\}.
\end{eqnarray}

\begin{eqnarray}
\fl
d_Q(y) =  {d_H\; y}
\Bigg\{ 1 - {q_0\over2} {y} 
+{1\over12}\left[3-2q_0+12q_0^2-4(j_0+ \Omega_0)\right] y^2
+ O(y^3) \Bigg\}.
\end{eqnarray}

\begin{eqnarray}
\fl
d_A(y) =  {d_H\; y}
\Bigg\{ 1 - {1\over2}\left[1+q_0\right] {y} 
+{1\over6}\left[q_0+3q_0^2-(j_0+\Omega_0)\right] y^2
+ O(y^3) \Bigg\}.
\end{eqnarray}
Note that in terms of the $y$ variable it is the ``deceleration distance'' $d_Q$ that has the deceleration parameter $q_0$ appearing in the simplest manner. Similarly, the ``logarithmic in distance'' Hubble relations are:
\begin{eqnarray}
\fl
\ln[d_L/(y \hbox{ Mpc})] &=& {\ln10\over5} [\mu_D - 25] - \ln y
\nonumber\\
\fl
&=& \ln(d_H/\hbox{Mpc})
\nonumber\\
\fl
&&
 -{1\over2}\left[-3+q_0\right] {y} 
+{1\over24}\left[21-2q_0+9q_0^2-4(j_0+\Omega_0)\right] y^2
+ O(y^3).
\end{eqnarray}

\begin{eqnarray}
\fl
\ln[d_F/(y \hbox{ Mpc})] &=& {\ln10\over5} [\mu_D - 25] - \ln y + {1\over2} \ln(1-y)
\nonumber\\
\fl
&=& \ln(d_H/\hbox{Mpc})
\nonumber\\
\fl
&&
 - {1\over2}\left[-2+q_0\right]  {y} 
+{1\over24}\left[15-2q_0+9q_0^2-4(j_0+\Omega_0)\right] y^2
+ O(y^3).
\end{eqnarray}

\begin{eqnarray}
\fl
\ln[d_P/(y \hbox{ Mpc})] &=& {\ln10\over5} [\mu_D - 25] - \ln y +  \ln(1-y)
\nonumber\\
\fl
&=& \ln(d_H/\hbox{Mpc})
\nonumber\\
\fl
&&
 - {1\over2}\left[-1+q_0\right]  {y} 
+{1\over24}\left[9-2q_0+9q_0^2-4(j_0+\Omega_0)\right] y^2
+ O(y^3).
\end{eqnarray}

\begin{eqnarray}
\fl
\ln[d_Q/(y \hbox{ Mpc})] &=& {\ln10\over5} [\mu_D - 25] - \ln y + {3\over2} \ln(1-y)
\nonumber\\
\fl
&=& \ln(d_H/\hbox{Mpc})
\nonumber\\
\fl
&&
 - {1\over2}q_0\,  {y} 
+{1\over24}\left[3-2q_0+9q_0^2-4(j_0+\Omega_0)\right] y^2
+ O(y^3).
\end{eqnarray}

\begin{eqnarray}
\fl
\ln[d_A/(y \hbox{ Mpc})] &=& {\ln10\over5} [\mu_D - 25] - \ln y + 2 \ln(1-y)
\nonumber\\
\fl
&=& \ln(d_H/\hbox{Mpc})
\nonumber\\
\fl
&&
 - {1\over2}\left[ 1 + q_0\right]  {y} 
+{1\over24}\left[-3-2q_0+9q_0^2-4(j_0+\Omega_0)\right] y^2
+ O(y^3).
\end{eqnarray}
Again note that the ``logarithmic in distance'' versions of the Hubble law are attractive in terms of maximizing the disentangling between Hubble distance, deceleration parameter, and jerk.
Now having a selection of Hubble laws on hand, for various definitions of cosmological distance and redshift, we can start to confront the observational data to see what it is capable of telling us. This is the basis of the analysis in~\cite{Hubble-arXiv}, which is well outside the scope of the current article.

\section{Conclusions}

Our main conclusions are threefold:
\begin{itemize}
\item 
The use of the $z$-redshift for $z>1$ is likely to lead to mathematical problems --- specifically any Taylor series in $z$ will be guaranteed to diverge for $z>1$, and so finite truncations will be poor approximations to the underlying physical function. This is not all that early in the evolution of the universe --- indeed many galaxies and supernovae are seen in the region $z\gtrsim 1$, so one ignores this issue at one's peril.

\item
The use of the $y$-redshift, where $y=z/(1+z)$, is very much to be encouraged for $z>1$ (corresponding to $y>1/2$). Taylor series in the   $y$-redshift are likely to be well behaved all the way back to the big bang (corresponding to $y=1$). 

\item
By combining the notions of $z$-redshift, $y$-redshift, and the many reasonably standard notions of ``cosmological distance'' that have appeared in the literature, it is possible to extract \emph{many} different versions of the Hubble law.  Which version of the Hubble law one chooses to adopt for any specific purpose will depend on the specific question being addressed. For instance, in~\cite{Hubble-arXiv} we have argued that plotting $\ln[d_F/z\hbox{ Mpc}]$ versus $z$, or $\ln[d_Q/y\hbox{ Mpc}]$ versus $y$, leads to a good visual test of the quality and robustness of the data underlying determinations of the deceleration parameter $q_0$.

\end{itemize}

\ack

This Research was supported by the Marsden Fund administered by the
Royal Society of New Zealand. CC was also supported by a Victoria University of Wellington Postgraduate scholarship.

\section*{References}
\addcontentsline{toc}{section}{References}


\end{document}